\documentclass[jcp,aip,reprint]{revtex4-1}
\usepackage{amsmath}
\usepackage{graphicx}
\usepackage{epstopdf}
\makeatletter
\g@addto@macro\bfseries{\boldmath}
\makeatother

\begin{document}

\title{The origin of unequal bond lengths in the $\tilde{\mbox{C}}$ $^1$B$_2$ state of SO$_2$: Signatures of high-lying potential energy surface crossings in the low-lying vibrational structure}
\author{G.\ Barratt Park}
\email{barratt.park@mpibpc.mpg.de}
\affiliation{Department of Chemistry, Massachusetts Institute of Technology, Cambridge, Massachusetts 02139}
\affiliation{Current address: Institute for Physical Chemistry, University of G\"{o}ttingen, Germany}
\author{Jun Jiang}
\affiliation{Department of Chemistry, Massachusetts Institute of Technology, Cambridge, Massachusetts 02139}
\author{Robert W.\ Field}
\affiliation{Department of Chemistry, Massachusetts Institute of Technology, Cambridge, Massachusetts 02139}

\begin{abstract}
The $\mathrm{\tilde{C}}$ $^1$B$_2$ state of SO$_2$ has a double-minimum potential in the antisymmetric stretch coordinate, such that the minimum energy geometry has nonequivalent SO bond lengths. The asymmetry in the potential energy surface is expressed as a staggering in the energy levels of the $\nu_3'$ progression. We have recently made the first observation of low-lying levels with odd quanta of  $v_3'$, which allows us---in the current work---to characterize the origins of the level staggering. Our work demonstrates the usefulness of low-lying vibrational level structure, where the character of the wavefunctions can be relatively easily understood, to extract information about dynamically important potential energy surface crossings that occur at much higher energy. The measured staggering pattern is consistent with a vibronic coupling model for the double-minimum, which involves direct coupling to the bound 2\,$^1$A$_1$ state and indirect coupling with the repulsive 3\,$^1$A$_1$ state. The degree of staggering in the $\nu_3'$ levels increases with quanta of bending excitation, which is consistent with the approach along the $\mathrm{\tilde{C}}$ state potential energy surface to a conical intersection with the 2\,$^1$A$_1$ surface at a bond angle of $\sim$145$^{\circ}$. 
\end{abstract}

\maketitle

\section{Introduction}
The $\tilde{\mbox{C}}$ $^1$B$_2$ state of SO$_2$ has in recent years attracted considerable attention because of its role in SO$_2$ photodissociation in the atmosphere.
\cite{Kawasaki1982377,KanamoriSO2TripletDissocMech,Effenhauser1990311,19912792,Becker_SO2_phofex,Becker_SO2_phofex2,okazaki:8752,Katagiri_SO2_photodissoc,ButlerSO2Emission,Sako1998571,Guo_SO2_3,Parsons2000499,SO2_dissoc__SOvibdist,Farquhar:2001fk,Gong2003493} 
However, earlier spectroscopy by Duchesne and Rosen\cite{duchesnerosenSO2}, Jones and Coon\cite{CoonSO2}, Brand and coworkers\cite{Brand_SO2_Cstate,BrandSO2MolPhys}, and by Hallin and Merer\cite{HallinThesis} focussed on the unusual low-lying vibrational structure below the dissociation limit, which was apparently the result of a distortion causing unequal SO bond lengths at the minimum-energy geometry. In the first paper of this series,\cite{SO2_IRUV_1} we report the first direct observations of $\mathrm{\tilde{C}}$-state levels with b$_2$ vibrational symmetry (odd quanta of $v_3'$), and in the second paper,\cite{SO2_IRUV_2} we report a new force field. This new information provides us with the opportunity to make a more precise characterization of the origins of level staggering than was previously possible. In the current paper (the third in the series), we present a vibronic model to explain the distortions in the low-lying vibrational structure of the $\mathrm{\tilde{C}}$ state, and we show that the vibronic (pseudo Jahn-Teller) distortion near equilibrium cannot be disentangled from the predissociation dynamics that occur at much higher energy. That is, we use low-lying vibrational energy level structure---where the wavefunctions can be relatively easily understood---to provide qualitative information about dynamical interactions that occur at much higher energies, where the level structure is less easy to interpret.

Ever since the initial spectroscopic investigations, the $\tilde{\mbox{C}}$ state of SO$_2$ has attracted a steady stream of theoretical attention. Mulliken first suggested that an unsymmetrical distortion of the S--O bond lengths might minimize antibonding in $\tilde{\mbox{C}}$-state SO$_2$,\cite{MullikenSO2AssymetryTheory} but Innes argued that the asymmetry in the potential is likely the result of vibronic interaction with a higher lying $^1$A$_1$ state.\cite{Innes19861} We believe Innes's argument to be the best explanation, but his analysis relies on an incorrect assignment of the $\nu_3'$ fundamental level by Ivanco and the derived parameters imply an unreasonably low energy for the perturbing electronic state. References \onlinecite{19912792,Nachtigall1999441,Bludsk2000607} report \textit{ab initio} calculations for the $\mathrm{\tilde{C}}$ state that reproduce the observed double-minimum potential energy surface. 
The low-lying vibrational structure of the $\mathrm{\tilde{C}}$ state has been calculated using an empirical potential obtained using an exact quantum mechanical Hamiltonian,\cite{Guo_SO2_3} and from a scaled \textit{ab initio} potential energy surface.\cite{Bludsk2000607} Both of these calculations are in excellent qualitative agreement with our observed staggering pattern, indicating that the asymmetry in the PES is well reproduced by the calculations. 
%

Due to the importance of SO$_2$ photodissociation in atmospheric chemistry, extensive experimental and theoretical work has focussed on the dissociative region of the $\mathrm{\tilde{C}}$-state PES above the dissociation limit, where the $\mathrm{\tilde{C}}$ (1\,$^1\mathrm{B}_2$) state undergoes a weakly avoided crossing with the 2\,$^1\mathrm{A}_1$ state and has a seam of intersection with the 1\,$^3\mathrm{A}_1$ state (3\,$^1\mathrm{A}'$ and 2\,$^3\mathrm{A}'$ in C$_{\mathrm{s}}$).
\cite{Okabe_SO2_Dissoc,Hui_SO2_dissoc,SO2ResonanceEmission,Kawasaki1982377,KanamoriSO2TripletDissocMech,Effenhauser1990311,
19912792,
Becker_SO2_phofex,Becker_SO2_phofex2,
Katagiri_SO2_photodissoc,
okazaki:8752,Sako1998571,ButlerSO2Emission,
Nachtigall1999441,Parsons2000499,Bludsk2000607,
SO2_dissoc__SOvibdist,Gong2003493,
GuoSO2IsotopomerSpectra,ZhangSO2ExcitedStates,Nanbu_SO2}
However, we are not aware of any detailed theoretical investigation of the $q_3$-mediated vibronic coupling between the $\mathrm{\tilde{C}}$ (1\,$^1\mathrm{B}_2$) level and the higher-lying 2\,$^1\mathrm{A}_1$ level in the diabatic basis. 

In the current work, we analyze the low-lying level structure of the $\mathrm{\tilde{C}}$ state in terms of a vibronic interaction model---inspired by the classic model of Innes\cite{Innes19861}---which indicates that the interaction of the $\mathrm{\tilde{C}}$ state with the quasi-bound 2\,$^1$A$_1$ state is probably influenced indirectly by the higher lying repulsive state, 3\,$^1$A$_1$. Our model is consistent with currently available theoretical results.\cite{19912792,Katagiri_SO2_photodissoc,Nachtigall1999441,Bludsk2000607,GuoSO2EmissionSpectra,PalmerSO2Cation,GuoSO2IsotopomerSpectra,ZhangSO2ExcitedStates,Nanbu_SO2} The success of our model demonstrates the use of low-lying features on the potential energy surface to obtain qualitative information about dynamics that emerge at much higher energies. This is an advantageous strategy in polyatomic molecules, because within a given electronic state, the complexity of the vibrational wavefunctions increases rapidly with energy. Low-lying vibrational fundamentals and overtones of small polyatomic molecules are usually well resolved and are often---to a good approximation---well described by normal mode quantum numbers in the product basis of harmonic oscillators. At high quanta of vibrational excitation, however, the vibrational eigenstates can usually only be described using a complicated linear combination of basis states, due to the increasing density of interacting basis states, leading ultimately to dynamics dominated by rapid intramolecular vibrational redistribution. In the $\mathrm{\tilde{C}}$ state of SO$_2$, the singlet avoided crossing occurs at $\sim$8000 cm$^{-1}$ above the $\mathrm{\tilde{C}}$-state origin, where a detailed interpretation of the vibrational level structure is not yet possible. However, the avoided crossing is related to the asymmetry near equilibrium, because both phenomena arise due to interactions among the same set of electronic states. Therefore, we can use the low-lying vibrational structure to extract qualitative information about higher-lying surface crossings.

\section{Vibrational level structure}

\begin{figure}
\centering
\includegraphics[width=\linewidth]{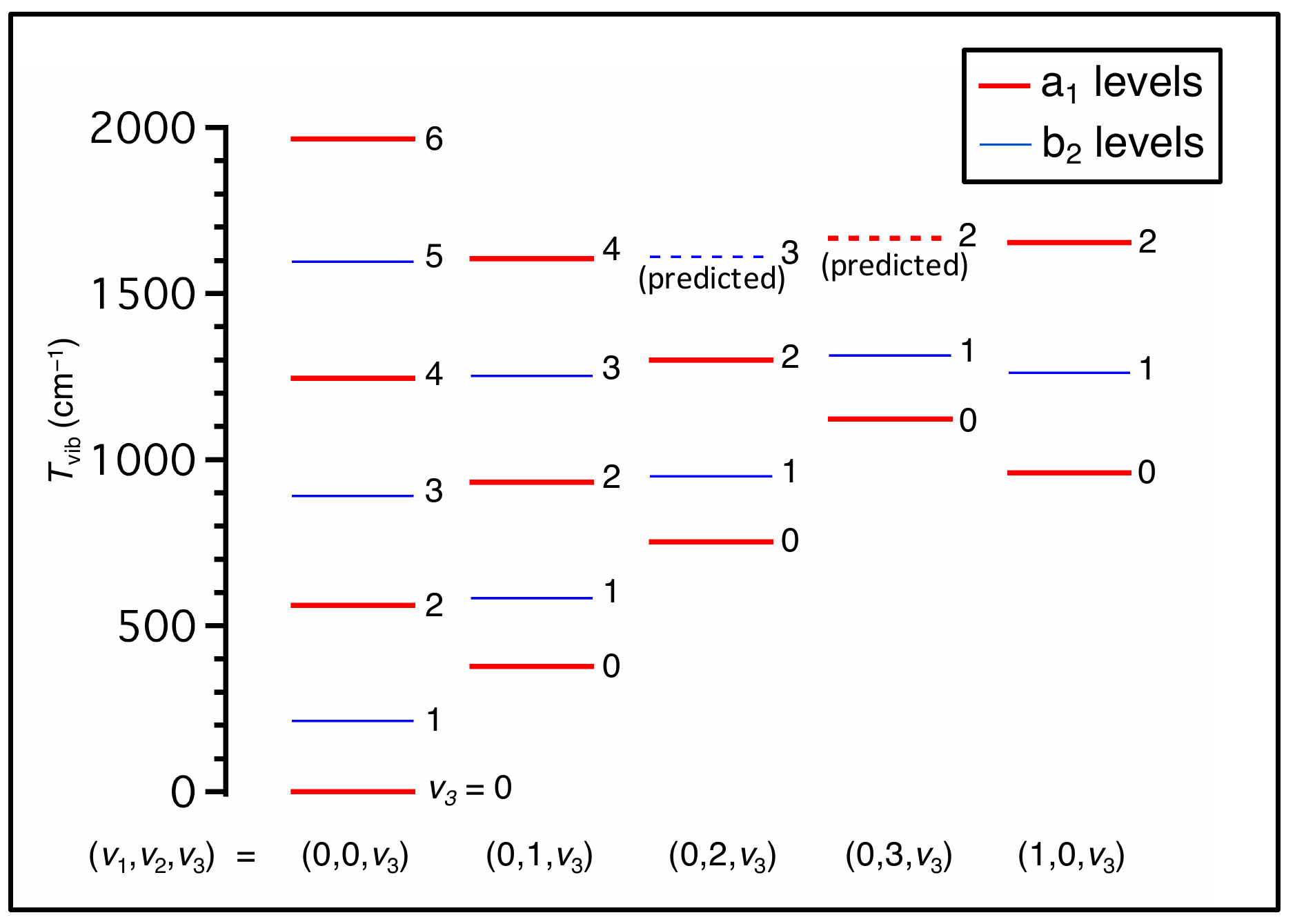}   \\
\caption{The low-lying vibrational level structure of the $\mathrm{\tilde{C}}$ state of SO$_2$ is shown, arranged as progressions in $\nu_3$.
}
\label{levelsfigure}
\end{figure}

\begin{figure}
\centering
\includegraphics[width=\linewidth]{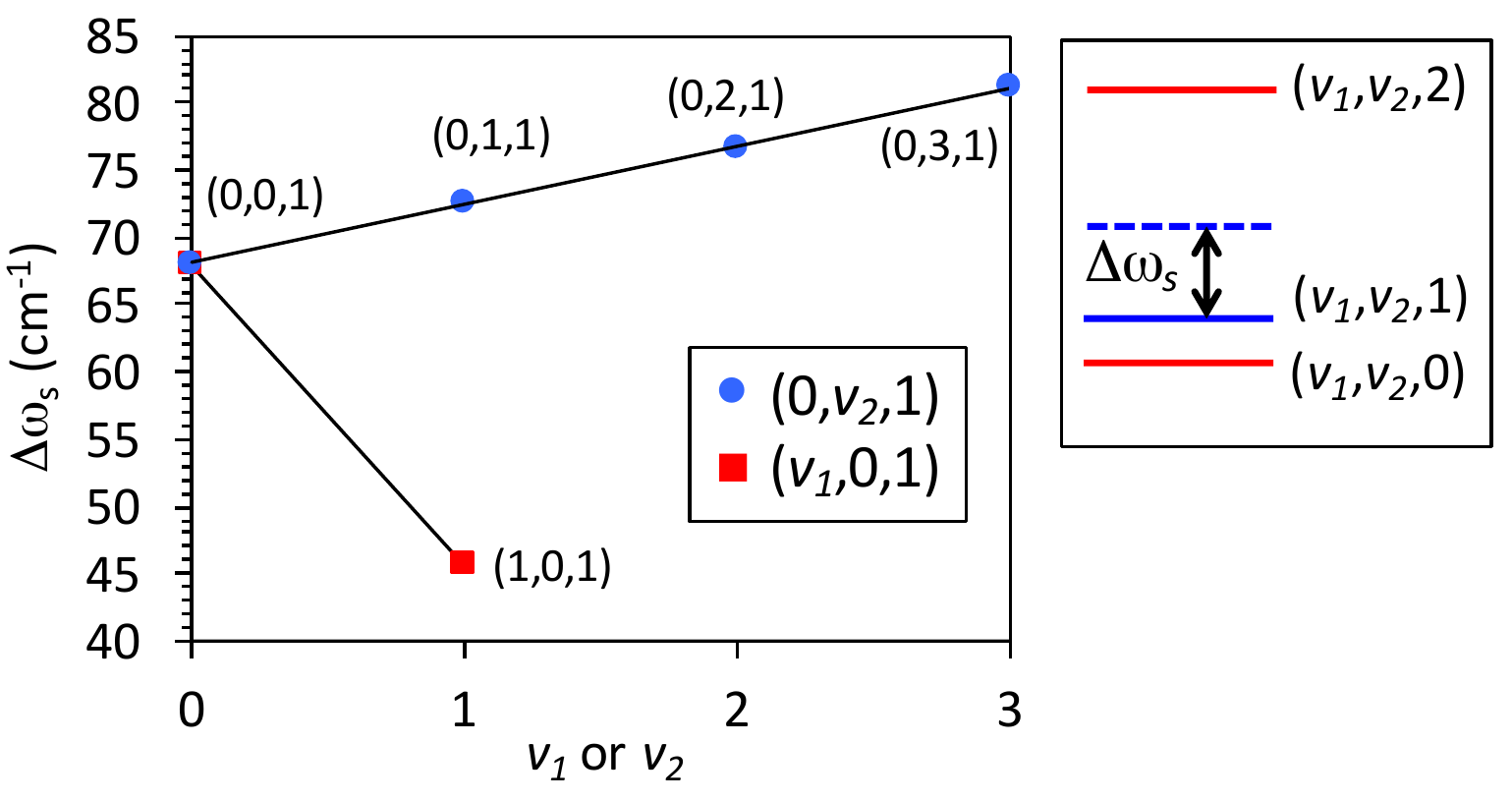}   \\
\caption{The staggering parameter, $\Delta\omega_s$, defined in Eq.\ (\ref{staggeringparameter}), is plotted as a function of $v_1$ and $v_2$. The parameter, shown schematically in the right panel of the figure, is related to the effective barrier height at the C$_{\mathrm{2v}}$ geometry. It increases linearly with $v_2$ as the $\mathrm{\tilde{C}}$-state PES approaches a conical intersection with the 2\,$^1\mathrm{A}_1$ potential at a bending angle of $\sim$145$^{\circ}$, which is consistent with a vibronic model for the double-well potential. 
}
\label{staggeringparamfig}
\end{figure}

The observed vibrational origins in the SO$_2$ $\mathrm{\tilde{C}}$ state up to 1600 cm$^{-1}$ above the $\mathrm{\tilde{C}}$(0,0,0) zero-point level are given in Tables VII and VIII of Ref.\ \onlinecite{SO2_IRUV_1}. In Fig.\ \ref{levelsfigure}, the energy level patterns are plotted for progressions in $v_3$. Due to the low barrier at the C$_{2\mathrm{v}}$ geometry, levels with a single quantum of $\nu_3$ are significantly depressed in frequency, but the magnitude of the odd/even level staggering decreases rapidly with increasing $v_3$, as the vibrational energy becomes large relative to the $\sim$100 cm$^{-1}$ barrier. We can define a parameter to characterize the degree of $\nu_3$ staggering as a function of the other vibrational quanta, $v_1$ and $v_2$:
\begin{equation}\label{staggeringparameter}
\Delta\omega_s(v_1,v_2)=\frac{T(v_1,v_2,0)+T(v_1,v_2,2)}{2}-T(v_1,v_2,1),
\end{equation}
where $T$ denotes the vibrational term energy and the notation $(v_1,v_2,v_3)$ is used for the vibrational quantum numbers. Equation \ref{staggeringparameter} gives the energy by which the \emph{expected} harmonic energy of $(v_1,v_2,1)$---which would be halfway between $(v_1,v_2,0)$ and $(v_1,v_2,2)$---is higher than the \emph{observed} energy of $(v_1,v_2,1)$, see inset of Figure \ref{staggeringparamfig}. A larger value of $\Delta\omega_s$ indicates an increased amount of staggering and a higher effective barrier height. 
The value of $\Delta\omega_s$ is plotted as a function of $v_1$ and $v_2$ in Figure \ref{staggeringparamfig}. The value of $\Delta\omega_s$ increases linearly with $v_2$ but decreases when one quantum of $v_1$ is added. As we will discuss Sec.\ \ref{DiscusionA}, the increase in $\Delta\omega_s$ with $v_2$ is consistent with a vibronic coupling model for the double-well potential, in which the asymmetry results from $q_3$-mediated interaction between the diabatic 1$^1$B$_2$ ($\mathrm{\tilde{C}}$) state and the 2\,$^1$A$_1$ state. 

\section{Interaction of the $\mathrm{\tilde{C}}$ state with 2\,$^1$A$_1$}\label{DiscusionA}
The avoided crossing between the 1$^1$B$_2$ ($\mathrm{\tilde{C}}$) and 2\,$^1$A$_1$ states has been extensively investigated at C$_\mathrm{s}$ geometries along the $\mathrm{SO}_2(\mathrm{\tilde{C}})\rightarrow\mathrm{SO}+\mathrm{O}$ photodissociation pathway.\cite{19912792,Katagiri_SO2_photodissoc,ButlerSO2Emission,Parsons2000499,GuoSO2EmissionSpectra,SO2_dissoc__SOvibdist,Bludsk2000607} The $\mathrm{\tilde{C}}$ state correlates diabatically to the excited singlet $\mathrm{SO}(^1\Delta)+\mathrm{O}(^1\mathrm{D})$ photodissociation products. However, the higher-lying 2\,$^3\mathrm{A}'$ and 3\,$^1\mathrm{A}'$ (1\,$^3$A$_1$ and 2\,$^1$A$_1$ in C$_{\mathrm{2v}}$) states both appear to correlate to the ground state triplet $\mathrm{SO}(^3\Sigma^-)+\mathrm{O}(^3\mathrm{P})$ product channel at geometries along the dissociation path. There is evidence for coupling of the $\mathrm{\tilde{C}}$ state to both the triplet and singlet dissociative states,\cite{KanamoriSO2TripletDissocMech,SO2_dissoc__SOvibdist} and both mechanisms probably contribute at different energies to the photodissociation of $\mathrm{\tilde{C}}$ state SO$_2$ to triplet products. (Coupling to the $\mathrm{\tilde{X}}$-state continuum is also believed to be an important mechanism.)\cite{Katagiri_SO2_photodissoc} However, to our knowledge, a full dimensional PES for the interacting $\mathrm{\tilde{C}}$ (1$^1$B$_2$) and $\mathrm{\tilde{D}}$ (2\,$^1$A$_1$) states has not been calculated, and in particular the interaction in the vicinity of the $\mathrm{\tilde{C}}$-state equilibrium has not received a thorough theoretical investigation, despite the suggestion by Innes that the double-well potential of the $\mathrm{\tilde{C}}$ state could arise from vibronic coupling to a \emph{bound} state of $^1$A$_1$ symmetry.\cite{Innes19861} (Vibronic coupling to a \textit{repulsive} diabat would \emph{not} yield a double-well potential surface.) 

\begin{figure}
\centering
\includegraphics[width=\linewidth]{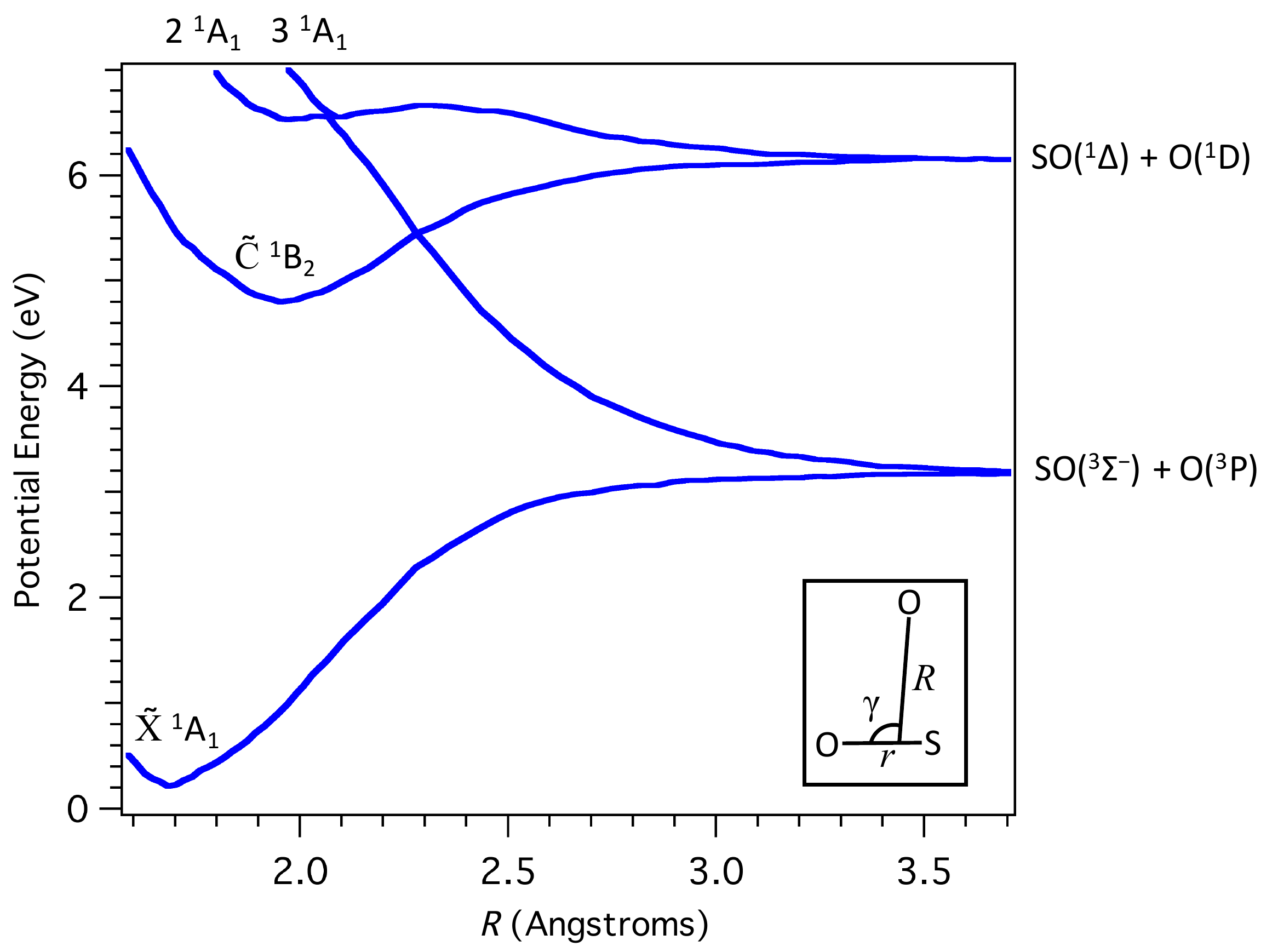}   \\
\caption{
The complete active space self-consistent field (CASSCF) potential energy curves---obtained from Ref.\ \onlinecite{Katagiri_SO2_photodissoc}---of four low-lying excited states of SO$_2$ are plotted as a function of the dissociative Jacobi coordinate, $R$. Values of the other Jacobi coordinates, defined in the inset, are fixed ($\gamma=120^{\circ}$ and $r=1.431$ \AA). To simplify the presentation, the electronic states not relevant to the current discussion are omitted from the figure. In this calculation, it is evident that the diabatic 3\,$^1$A$_1$ state is repulsive, correlating with the ground state $\mathrm{SO}(^3\Sigma^-)+\mathrm{O}(^3\mathrm{P})$ product channel, but the diabatic 2\,$^1$A$_1$ state appears to be bound, correlating to a higher-lying product channel. In this figure, all excited states appear less energetically stable than in reality because the one-dimensional potential energy slice does not sample the equilibrium geometry of each state. 
}
\label{KatagiriFig}
\end{figure}

Although the discussion in Refs.\ \onlinecite{19912792,Katagiri_SO2_photodissoc} mostly focusses on the dissociative region of the PES, there is evidence in the calculations that the 2\,$^1$A$_1$ state is bound at the C$_{\mathrm{2v}}$ geometry and becomes dissociative (towards the ground state product channel) as a result of interaction with the repulsive 3\,$^1$A$_1$ state. Figure \ref{KatagiriFig}, which is derived from Fig.\ 6 of Ref.\ \onlinecite{Katagiri_SO2_photodissoc}, displays a calculated one-dimensional slice through the potential energy surfaces of low-lying electronic states of SO$_2$ along the dissociation coordinate. Ray \textit{et al.}\cite{ButlerSO2Emission} calculated oscillator strengths for vertical electronic excitation from the ground state to the $\mathrm{\tilde{C}}$(1$^1$B$_2$), 2\,$^1$A$_1$, and 3\,$^1$A$_1$ states (See Table I of Ref.\ \onlinecite{ButlerSO2Emission}.) The calculated result was $f=0.1336$ for the 1$^1$B$_2$ ($\mathrm{\tilde{C}}$) state; $f=0.0290$ for the 2\,$^1$A$_1$ state; and $f=0.1625$ for the 3\,$^1$A$_1$ state, where $f$ is the oscillator strength. Therefore, the 3\,$^1$A$_1\leftarrow\mathrm{\tilde{X}}$ transition strength is expected to be comparable to that of the $\mathrm{\tilde{C}}\leftarrow\mathrm{\tilde{X}}$ transition, but the 2\,$^1$A$_1\leftarrow\mathrm{\tilde{X}}$ transition is expected to be weaker by almost an order of magnitude. This result would explain why $q_3$-mediated vibronic interaction of the $\mathrm{\tilde{C}}$ state with 2\,$^1$A$_1$ near equilibrium geometry could be strong enough to result in a double-minimum potential, but does not lend one-photon vibronically allowed intensity into b$_2$ vibrational levels, because the oscillator strength of the 2\,$^1$A$_1\leftarrow\mathrm{\tilde{X}}$ transition is weak. On the other hand, near the avoided crossing along the dissociation coordinate, 3\,$^1$A$'$ correlates diabatically to the 3\,$^1$A$_1$ (4$^1$A$'$ in C$_{\mathrm{s}}$) state, which has a much larger oscillator strength. As noted by Ray \textit{et al.},\cite{ButlerSO2Emission} this may be one of the reasons why dispersed fluorescence experiments from energies near the avoided crossing give rise to intensity into \emph{both} a$_1$ and b$_2$ vibrational levels, although we believe Coriolis interactions probably also contribute, given the high vibrational level density in this region. 

Very little theoretical work has been done to determine the equilibrium structure of the $\mathrm{\tilde{D}}$ 2\,$^1$A$_1$ state at bound geometries. Nevertheless, it appears that $q_3$-mediated vibronic interaction with the 2\,$^1$A$_1$ state may be responsible for \emph{both} the double well potential of the $\mathrm{\tilde{C}}$ state near equilibrium \emph{and} the avoided crossing that causes the $\mathrm{\tilde{C}}$-state to dissociate adiabatically to the ground state product channel. Thus, this is a case where a detailed understanding of the PES near equilibrium is highly relevant to dissociation processes that take place far from equilibrium, because both are influenced by vibronic coupling involving the same higher-lying electronic states. 

\subsection{One dimensional vibronic coupling model}
Following Innes,\cite{Innes19861} we begin our analysis of vibronic coupling with a simple one-dimensional model involving two electronic states. We assume that both electronic states, in zero order, behave like simple harmonic oscillators in the antisymmetric stretch coordinate, but we allow the harmonic oscillators to have different frequencies. We write our model Hamiltonian in the diabatic basis of separable vibration-electronic states, $|\psi_n^{\mathrm{vib}}(q_3)\rangle|\psi_j^{\mathrm{el}}\rangle$, where $|\psi_n^{\mathrm{vib}}(q_3)\rangle$ are the harmonic oscillator basis states with harmonic frequency $\omega_3$, and $|\psi_j^{\mathrm{el}}\rangle$ represents the lower and upper interacting electronic states with $j=a$ or $b$, respectively:
\begin{equation}
\begin{aligned}[b] 
\label{InnesH}
\mathbf{H}&=\mathbf{H}_0+\mathbf{H'} \\
\mathbf{H}_0&=\omega_3\left(\hat{N}+\frac{1}{2}\right)
|\psi_a^{\mathrm{el}}\rangle\langle\psi_a^{\mathrm{el}}|
\\&+ \left[\frac{1}{2}\,\omega_3'\!\left( \frac{\omega_3'}{\omega_3}+\frac{\omega_3}{\omega_3'}\right)
\left(\hat{N}+\frac{1}{2} \right)+D_{ab}\right]
|\psi_b^{\mathrm{el}}\rangle\langle\psi_b^{\mathrm{el}}| \\
\mathbf{H}'&=\lambda_{ab}\boldsymbol{q}_3\left(  |\psi_a^{\mathrm{el}}\rangle\langle\psi_b^{\mathrm{el}}|+|\psi_b^{el}\rangle\langle\psi_a^{el}|  \right)
\\&+ \frac{1}{4}\, \omega_3'\! \left( \frac{\omega_3'}{\omega_3}-\frac{\omega_3}{\omega_3'} \right)(\hat{a}_3\hat{a}_3+\hat{a}_3^{\dagger}\hat{a}_3^{\dagger})
|\psi_b^{\mathrm{el}}\rangle\langle\psi_b^{\mathrm{el}}|,
\end{aligned}
\end{equation}
where $D_{ab}$ gives the energy spacing between the electronic states, $\lambda_{ab}$ is a vibronic coupling constant, and $\omega_3'$ is the harmonic frequency of the upper electronic state. $\hat{N}$ and $\hat{a}$ represent the quantum harmonic oscillator number operator and annihilation operator, respectively. Diagonal matrix elements are given by $\mathbf{H}_0$. The first term in $\mathbf{H}'$ gives rise to $\Delta v_3=\pm1$ matrix elements that couple levels of different electronic states, and the second term gives rise to $\Delta v_3=\pm2$ matrix elements in the excited electronic state, which arise from the rescaling of the dimensionless $\boldsymbol{p}$ and $\boldsymbol{q}$ operators for the vibrational frequency of the upper state. In other words, this term must be included because the excited state, with harmonic frequency $\omega_3'$, is being described in the basis of a harmonic oscillator of a different frequency, $\omega_3$. Note that this last term vanishes for the simplifying case when $\omega_3'=\omega_3$, and the $\frac{1}{2}(\omega_3'/\omega_3+\omega_3/\omega_3')$ scaling factor in $\mathbf{H}_0$ becomes unity. The Hamiltonian in Eq.\ \ref{InnesH} gives rise to matrix elements of the form
\begin{subequations}
\begin{align}
\langle\psi_a^{\mathrm{el}}\psi_n^{\mathrm{vib}}|\mathbf{H}|\psi_a^{\mathrm{el}}\psi_n^{\mathrm{vib}}\rangle
&=\omega_3\left(n+\frac{1}{2}\right) \\
\langle\psi_b^{\mathrm{el}}\psi_n^{\mathrm{vib}}|\mathbf{H}|\psi_b^{\mathrm{el}}\psi_n^{\mathrm{vib}}\rangle \nonumber
\\=\frac{1}{2}\omega_3'    \left(\frac{\omega_3'}{\omega_3}\right. + &\left.\frac{\omega_3}{\omega_3'}\right)     \left(n+\frac{1}{2} \right)  +D_{ab} \\
\langle\psi_b^{\mathrm{el}}\psi_{n\pm2}^{\mathrm{vib}}|\mathbf{H}|\psi_b^{\mathrm{el}}\psi_n^{\mathrm{vib}}\rangle \nonumber
\\= \frac{1}{4} \omega_3'   \left( \frac{\omega_3'}{\omega_3}\right. -  &\left. \frac{\omega_3}{\omega_3'}  \right)   \sqrt{(n\pm1)(n+1\pm1)} \\
\langle\psi_a^{\mathrm{el}}\psi_{n\pm1}^{\mathrm{vib}}|\mathbf{H}|\psi_b^{\mathrm{el}}\psi_n^{\mathrm{vib}}\rangle
&=\lambda_{ab}\sqrt{\frac{1}{2}\left( n+\frac{1}{2}\pm\frac{1}{2}  \right)},\label{needsCC}
\end{align}
\end{subequations}
as well as the complex conjugate of Eq.\ \ref{needsCC}.

We fit the frequencies of the $(0,0,v_3)$ progression to the Hamiltonian in Eq.\ \ref{InnesH} by truncating and diagonalizing the matrix. In order to ensure a physically realistic result, we constrain $D_{ab}$ to the calculated difference in energy for vertical excitation of SO$_2$ to the $\mathrm{\tilde{C}}$ 1$^1$B$_2$ and the 2\,$^1$A$_1$ states from Ref.\ \onlinecite{Katagiri_SO2_photodissoc}, and we constrained $\omega_3$ to a ``normal'' value (we use the $\omega_3$ frequency in the ground electronic state.) The results are shown in Table \ref{InnesFit}. The $(0,0,v_3)$ progression is qualitatively reproduced by the model, although the fit is far from spectroscopically accurate. It is possible to achieve much better agreement (rms error $1.87$ cm$^{-1}$) by removing the constraints on $D_{ab}$ and $\omega_3$, but the best fit values are much lower and higher, respectively, than our physically reasonable estimate. The simplistic one-dimensional vibronic coupling model ignores all other sources of anharmonicity, and is therefore not expected to give quantitative results. The best fit parameters underestimate the degree of level staggering, so it is possible that the vibronic interaction parameter $\lambda_{ab}=2297$ cm$^{-1}$ is too low. However the ability of the model to qualitatively reproduce the $(0,0,v_3)$ level structure is good evidence for the presence of vibronic coupling, as first suggested by Innes.\cite{Innes19861}

\begin{table}
\caption{
Results of a fit of the measured $(0,0,v_3)$ vibrational term energies to the one-dimensional vibronic coupling model of Eq.\ \ref{InnesH}. The values of $D_{ab}$ and $\omega_3$ were constrained. The model is qualitative in nature, and we estimate the uncertainty in the parameters to be on the order of 30\%. All values are given in cm$^{-1}$ units.
 \label{InnesFit}} 
\vspace{3 pt}
\centering
\begin{tabular}{rrr} \toprule
Level & $T_{\mathrm{vib}}(\mathrm{exp})$ & $T_{\mathrm{vib}}(\mathrm{fit})$ \\ 
\colrule
(0,0,1) & 212.575   & 227.76 \\
(0,0,2) & 561.232   & 544.82 \\
(0,0,3) & 890.939   & 886.76 \\
(0,0,4) & 1245.469 & 1249.75 \\
(0,0,5) & 1595.794 & 1626.53 \\
\colrule
Parameters: & $D_{ab}=14760$                 & $\omega_3=1362$ \\
                       &$\lambda_{ab}=2297$            &$\omega_3'=451.7$ \\
\botrule                         
\end{tabular}
\end{table}

At first glance, the extremely low harmonic frequency, $\omega_3'$, obtained for the perturbing state might appear alarming. However, this low value of $\omega_3'$ is crucial to the success of the model, and the explanation is straightforward. As mentioned in Section \ref{DiscusionA}, calculations suggest that the 2\,$^1$A$_1$ state is only quasi-bound. An avoided crossing with 3\,$^1$A$_1$, which lies only $\sim$0.5 eV higher in energy, causes 2\,$^1$A$_1$ to become dissociative, correlating to the ground state $\mathrm{SO}(^3\Sigma^-)+\mathrm{O}(^3\mathrm{P})$ dissociation channel. Such an interaction could dramatically decrease the effective $\omega_3'$ harmonic frequency of 2\,$^1$A$_1$, because of mode softening along the dissociative coordinate. Thus, the same set of interactions that contribute to photodissociation of SO$_2$ via singlet vibronic coupling at around 48,000 cm$^{-1}$ also appear to be the direct cause of the unusual vibrational structure near the bottom of the $\mathrm{\tilde{C}}$-state potential energy surface at around 43,000 cm$^{-1}$. This underscores the importance of understanding the low-lying vibrational level structure, where the spectroscopic information is comparatively simple, yet mechanistic information can be gleaned about dynamics that appear at much higher energy. 

\begin{figure*}
\centering
\includegraphics[width=\linewidth]{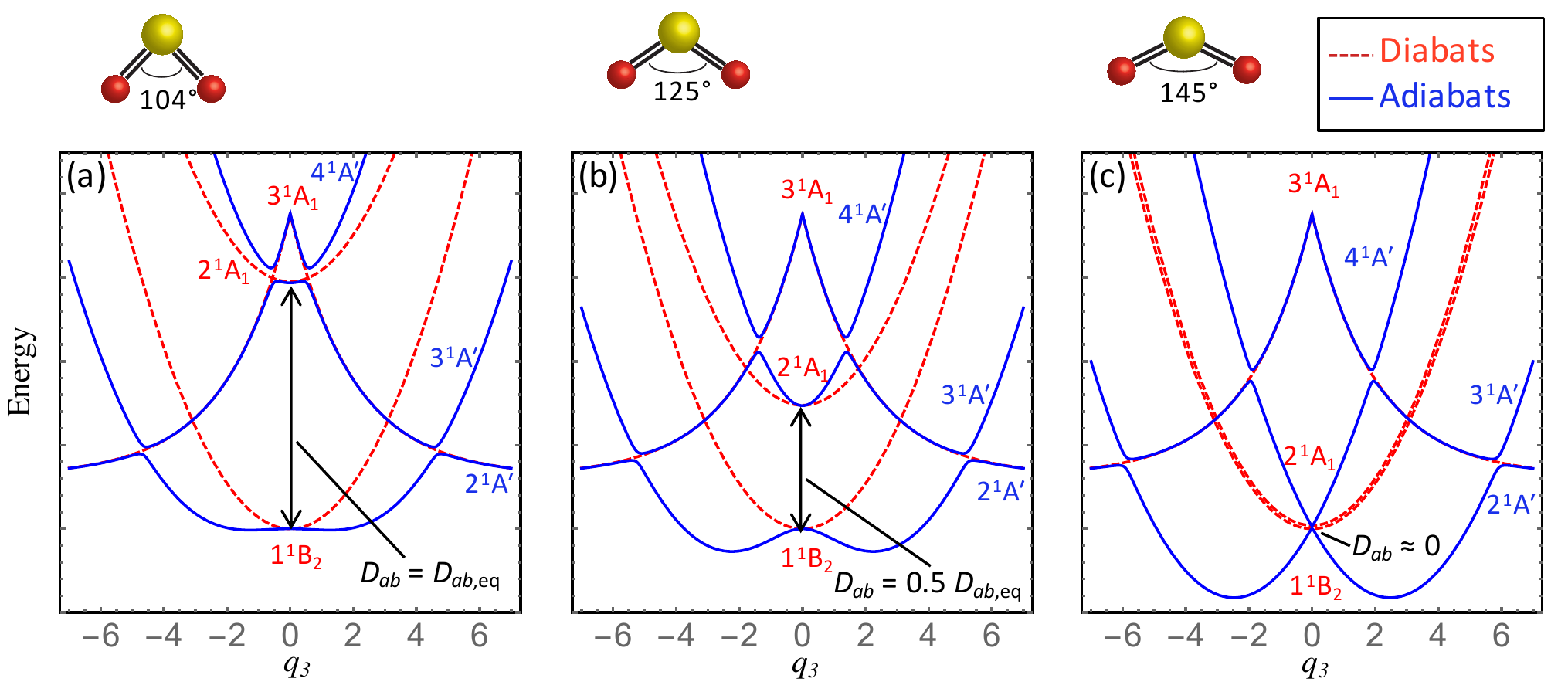}   
\caption{A toy one-dimensional model for the $q_3$-mediated vibronic interaction between the 1$^1\mathrm{B}_2$ ($\mathrm{\tilde{C}}$) state and the 2$^1\mathrm{A}_1$ state is illustrated. The model is calculated from the Hamiltonian in Eq.\ (\ref{ToyEquation}) and is shown schematically as a function of $D_{ab}$, which gives the separation between the two bound diabatic states. Interaction with the dissociative 4$^1\mathrm{A}'$ state is also included. The values of the parameters, in cm$^{-1}$, are $D_{ab,\mathrm{eq}}=14760$, $D_{ac}=18792$, $D_0=3152$, $\lambda_{ab}=3400$, $\lambda_{bc}=500$, $\omega_3=1350$, and the unitless parameter $l=2$. The two bound potential energy surfaces cross via a conical intersection that occurs at a bond angle of $\sim$145$^{\circ}$, but the crossing is avoided at C$_{\mathrm{s}}$ geometries. Therefore, as the bond angle increases, the energy denominator for the vibronic interaction decreases and the effective barrier in the $\mathrm{\tilde{C}}$-state adiabatic potential energy surface increases, consistent with the observations shown in Figure \ref{staggeringparamfig}. 
}
\label{ConicalIntersectionFig}
\end{figure*}

To illustrate more explicitly the three-state system that gives rise to the structure near the bottom of the $\mathrm{\tilde{C}}$-state potential energy surface, we construct a toy one-dimensional model for the adiabatic potential energy curves of the bound $\mathrm{\tilde{C}}$ 1$^1$B$_2$ and 2\,$^1$A$_1$ states and the higher-lying repulsive 3\,$^1$A$_1$ ($4^1$A$'$) state. In the diabatic basis, the toy Hamiltonian is
\begin{equation}\label{ToyEquation}
\mathbf{H}=
\begin{pmatrix}
M_a     & V_{ab} & 0 \\
V_{ab} & M_b     & V_{bc} \\
0           & V_{bc} & M_c
\end{pmatrix},
\end{equation}
where the matrix elements have the form
\begin{equation*}
\begin{aligned}
V_{ab}=\lambda_{ab}q_3, && V_{bc}=\lambda_{bc}, \\
M_a=\frac{1}{2}\omega_3q_3^2, && M_b=\frac{1}{2}\omega_3q_3^2+D_{ab}, \\
M_c=\ \span\span(D_{ac}-D_0)\exp{\left(-|q_3/l|\right)}+D_0.
\end{aligned}
\end{equation*}
We assume that the $V_{ab}$ interaction is vibronic in nature since it couples states of different electronic symmetry in C$_{\mathrm{2v}}$ ($^1$A$_1$ to $^1$B$_2$), but the $V_{bc}$ interaction is assumed to be vibrationally independent since it couples states of the same ($^1$A$_1$) electronic symmetry. The parameter $D_0$ is the energy of the ground state dissociation channel, and $D_{ab}$ and $D_{ac}$ characterize the energy spacing between the $\mathrm{\tilde{C}}$ state and the two higher lying electronic states. The $l$ parameter is the characteristic decay length of the repulsive state.  The one-dimensional diabats and adiabats of the toy model, obtained with `best guess' values of the parameters, are plotted in Figure \ref{ConicalIntersectionFig}a. In Sec.\ \ref{conintsec}, we will use this qualitative figure as a starting point to extend the discussion of the vibronic interaction to other vibrational coordinates. 

\subsection{Evidence for increased effective barrier height along the approach to conical intersection}\label{conintsec}
The $\mathrm{\tilde{C}}$ 1$^1$B$_2$ and $\mathrm{\tilde{D}}$ 2\,$^1$A$_1$ states belong to different symmetry species in C$_{\mathrm{2v}}$, but they both correlate to $^1$A$'$ in C$_{\mathrm{s}}$ geometries. Therefore, although the crossing is avoided at C$_{\mathrm{s}}$ geometries, the levels may cross in C$_{\mathrm{2v}}$ geometries, resulting in a seam of conical intersection. Theoretical investigations\cite{Katagiri_SO2_photodissoc,Bludsk2000607} have reported the lowest seam of intersection to occur at bond angles between 145--150$^{\circ}$ in C$_{\mathrm{2v}}$ for bond lengths near the equilibrium value. This is a much wider bond angle than the $\sim$104$^{\circ}$ equilibrium bond angle of the $\mathrm{\tilde{C}}$ state. If the double-minimum potential of the $\mathrm{\tilde{C}}$ state is caused by $q_3$-mediated vibronic interactions with 2\,$^1$A$_1$ around the C$_{\mathrm{2v}}$ equilibrium, we expect the effect to become very strong at geometries near the conical intersection, since the energy denominator for the interaction vanishes at the conical intersection. As quanta of $v_2$ are added, the vibrational wavefunction has increased amplitude at wider bond angles, as indicated by the large negative value of the $\alpha_2^A$ rotation-vibration constant (i.e.\ the effective $A$ constant increases as the bond angle is widened towards linearity---see Table IX and Figure 6b of Ref.\ \onlinecite{SO2_IRUV_1}.) Therefore, it is highly likely that the increase in $\Delta\omega_s$ as a function of $v_2$ (Figure \ref{staggeringparamfig}) is a direct consequence of the approach to the seam of conical intersection. To illustrate this point, we calculate the toy model adiabatic potential energy curves from Eq.\ (\ref{ToyEquation}) with reduced values of the energy difference, $D_{ab}$. Figure \ref{ConicalIntersectionFig}(b) shows the result when $D_{ab}$ is equal to half of its equilibrium value ($104^{\circ}<\angle\mathrm{OSO}<145^{\circ}$), and Fig.\ \ref{ConicalIntersectionFig}(c) shows the result at the conical intersection ($\angle\mathrm{OSO}=145^{\circ}$), where $D_{ab}=0$.

Without more detailed knowledge of the 2\,$^1$A$_1$ potential energy surface, it is difficult to make a quantitative prediction of the expected trend in $\Delta\omega_s$ as a function of $v_2$, which results from the approach to conical intersection. However, we can estimate the trend by building a simple model. Our approach will be to approximate the vibrationally-averaged energy difference between the $\mathrm{\tilde{C}}$ $^1$B$_2$ and the 2\,$^1$A$_1$ surfaces as a function of $v_2$ bending quanta in the $\mathrm{\tilde{C}}$ state. We will then use the vibrationally averaged energy difference to calculate $\Delta\omega_s(v_2)$ from the one-dimensional vibronic model (Eq.\ (\ref{InnesH})).

According to the calculations in Refs.\ \onlinecite{19912792} and \onlinecite{Katagiri_SO2_photodissoc}, the 2\,$^1$A$_1$ state appears to have a wide equilibrium bond angle ($\sim$160$^{\circ}$), but a $\omega_2$ bending frequency similar to that of the $\mathrm{\tilde{C}}$ state. We therefore model the one-dimensional bending potential energy curves of the upper ($V_b$) and lower ($V_a$) states as
\begin{equation}\label{D1eff}
\begin{aligned}[b]
V_a(q_2)&=\frac{1}{2}\omega_2 q_2^2+\frac{1}{6}\phi_{222}q_2^3+\frac{1}{24}\phi_{2222}q_2^4 \\
V_b(q_2)&=\frac{1}{2}\omega_2 (q_2-\delta)^2 +D \\
\end{aligned}
\end{equation}
\begin{equation*}
\begin{aligned}
\omega_2&=392.28 &\phi_{222}&=-85.375 \\
\phi_{2222}&=8.276  &\delta&=3.0 & D=13059,
\end{aligned}
\end{equation*}
where values for $\omega_2$, $\phi_{222}$, and $\phi_{2222}$, in cm$^{-1}$, are taken from our $\mathrm{\tilde{C}}$-state force field fit reported in Part II of this series,\cite{SO2_IRUV_2} and $\delta$ gives the approximate equilibrium displacement of the excited 2\,$^1$A$_1$ state, $160^{\circ}-104^{\circ}=56^{\circ}$, in dimensionless normal mode coordinates obtained from the same force field. The value of $D$ (in cm$^{-1}$) was chosen in order to make the energy difference between the displaced potential energy curves match the value $D_{ab}=14760$ cm$^{-1}$ given in Table \ref{InnesFit} near the geometry of the $\mathrm{\tilde{C}}$-state equilibrium. We calculate the low-lying one-dimensional vibrational wavefunctions, $\psi_{v_2}(q_2)$, of $V_a$ using discrete variable representation, and we integrate to obtain the vibrationally averaged expectation value for the energy difference, 
\begin{equation}\label{DeffInt}
\langle D_{ab}(v_2)\rangle=\int_{-\infty}^{\infty}{\psi_{v_2}(q_2)\left[V_b(q_2)-V_a(q_2)\right]\psi_{v_2}(q_2)\, \mathrm{d}q_2}.
\end{equation}
The resulting values of $\langle D_{ab}(v_2)\rangle$ are then substituted into the vibronic coupling model (Eq.\ (\ref{InnesH})), in order to calculate the staggering parameter $\Delta\omega_s$, defined in Eq.\ (\ref{staggeringparameter}). The results are tabulated in Table \ref{ConicalIntersectionModel}. The model (Eq.\ (\ref{D1eff}--\ref{DeffInt})) predicts that the energy difference parameter $\langle D(v_2)\rangle$ decreases linearly by $\sim$130 cm$^{-1}$ per quantum of bend excitation. Although the parameters of our harmonic, one-dimensional vibronic model (Table \ref{InnesFit}), underestimate the staggering parameter, $\Delta\omega_s$, by approximately 23 cm$^{-1}$, the overall interaction model reproduces the observed trend in $\Delta\omega_s(0,v_2)$ very well. The model predicts a nearly linear increase in $\Delta\omega_s$ of 5.3 cm$^{-1}$ per quantum of $v_2$, whereas the experimentally determined trend is 4.4 cm$^{-1}$ per quantum. The experimental trend in $\Delta\omega_s(0,v_2)$ is thus consistent with the proposed vibronic interaction model, and further illustrates the capability of low-lying features on the potential energy surface to provide information about phenomena that occur at much higher energy. In this case, the trend in vibrational level staggering induced by a spectator mode ($\nu_2$) acts as an early warning signal that alerts us to the approach to a conical intersection as the geometry is displaced along that mode. 

We note that this type of effect, involving a totally symmetric spectator mode, is unique to \textit{pseudo} Jahn-Teller systems, where a vibronic interaction between \textit{non-degenerate} electronic states leads to a distorted minimum-energy configuration. In this type of system, the two electronic states are---in general---not degenerate, even at the symmetric configuration involving zero displacement along the non-totally symmetric coordinate, but may cross at a seam of conical intersection that occurs for particular displacements along the totally symmetric coordinates. In a \textit{true} Jahn-Teller system, involving degenerate zero-order electronic states, totally symmetric spectator mode effects are not expected to occur, because the electronic states are necessarily degenerate at \textit{any} configuration of the higher-symmetry point group of the zero-order states (i.e. at configurations involving arbitrary displacement along the totally symmetric coordinates, but zero displacement along the non-totally symmetric coordinate.)

\begin{table}
\caption{
The vibrationally-averaged electronic state separation, $\langle D_{ab}(v_2)\rangle$, from Eq.\ (\ref{D1eff}--\ref{DeffInt}), and the resulting value of the staggering parameter, $\Delta\omega_s(0,v_2)$, obtained from the vibronic coupling model (Eq.\ (\ref{InnesH})), with $D_{ab}=\langle D_{ab}(v_2)\rangle$ (Eq.\ (\ref{DeffInt})). The increase in $\Delta\omega_s(0,v_2)$ per quantum of $v_2$, $\Delta\Delta\omega_s(0,v_2)=\Delta\omega_s(0,v_2)-\Delta\omega_s(0,v_2-1)$, is also tabulated.  The experimentally-determined $\Delta\omega_s(0,v_2)$ and $\Delta\Delta\omega_s(0,v_2)$ values are listed for comparison. All energies are in cm$^{-1}$ units.
 \label{ConicalIntersectionModel}} 
\vspace{3 pt}
\centering
\resizebox{\linewidth}{!}{ 
\begin{tabular}{@{\extracolsep{6pt}}cccccc} \toprule
           & \multicolumn{3}{c}{Model} & \multicolumn{2}{c}{Expt.} \\
           \cline{2-4} \cline{5-6}
$v_2$ & $\langle D_{ab}(v_2)\rangle$ & $\Delta\omega_s(0,v_2)$ &$\Delta\Delta\omega_s(0,v_2)$& $\Delta\omega_s(0,v_2)$ &$\Delta\Delta\omega_s(0,v_2)$\\
\colrule
0          &14760                                             &44.66        &                   &68.02  &      \\
1          &14630                                             &49.51        &4.85            &72.52  &4.50\\
2          &14497                                             &54.85        &5.34            &76.65  &4.13\\
3          &14362                                             &60.68        &5.83            &81.23  &4.58\\
\botrule
\end{tabular}
}
\end{table}

Our results support our proposed vibronic coupling mechanism with the 2\,$^1$A$_1$ state and also provide predictions against which to test theoretical investigations. To our knowledge, calculation of a full dimensional PES for the interacting 1$^1$B$_2$ ($\mathrm{\tilde{C}}$) and 2\,$^1$A$_1$ states has not been performed, and the location of the conical intersection as a function of $q_1$ has not been investigated. However, if the decrease in $\Delta\omega_s(v_1,0)$ (see Fig.\ \ref{staggeringparamfig}) is influenced by the location of the seam of conical intersection in a similar manner as the trend in $\Delta\omega_s(0,v_2)$, this would suggest that at the $\mathrm{\tilde{C}}$ state equilibrium bond angle, the conical intersection occurs at \emph{shorter} than the effective C$_{\mathrm{2v}}$ equilibrium bond distance of 1.576 \AA. That is, as the effective bond lengths are increased, the strength of the vibronic interaction decreases, indicating an increase in the energy denominator for the vibronic interaction.

\section{Conclusions}

Our observations (reported in Part I of this series)\cite{SO2_IRUV_1} are consistent with a vibronic coupling model for the asymmetric equilibrium bonding structure, first proposed by Innes,\cite{Innes19861} in which the $\mathrm{\tilde{C}}$ state undergoes a $q_3$-mediated interaction with the (diabatically) bound 2\,$^1$A$_1$ state. The oscillator strength of the 2\,$^1$A$_1\leftarrow \mathrm{\tilde{X}}$ 1\,$^1$A$_1$ transition is calculated to be relatively weak at the equilibrium C$_{2v}$ geometry, which is consistent with the fact that no vibronically-allowed one-photon transitions to low-lying b$_2$ vibrational levels of the $\mathrm{\tilde{C}}$ state have been observed. As noted in Ref.\ \onlinecite{ButlerSO2Emission}, vibronically allowed transitions that violate the vibrational selection rules are plausible at higher energies near the avoided crossing of the $\mathrm{\tilde{C}}$-state with 2\,$^1$A$_1$ in the dissociative region, because 2\,$^1$A$_1$ probably borrows oscillator strength via an avoided crossing with the dissociative 3\,$^1$A$_1$ state. 

Using information from the low-lying vibrational levels of the $\mathrm{\tilde{C}}$-state, we are able to develop a picture that accounts for these three interacting electronic states. Our one-dimensional two-state vibronic model fails to reproduce the observed level pattern in the $(0,0,v_3)$ progression unless an anomalously low value of $\omega_3'$ is chosen for the upper state. This may suggest an indirect role that the repulsive 3\,$^1$A$_1$ state plays in shaping the adiabatic $\mathrm{\tilde{C}}$-state potential energy surface. Interaction of 2\,$^1$A$_1$ with 3\,$^1$A$_1$ may dramatically decrease the effective $\omega_3'$ frequency of 2\,$^1$A$_1$, giving rise to the low value of $\omega_3'$ in our fit model. The apparent involvement of 3\,$^1$A$_1$ in the observed level structure has profound implications for the photodissociation dynamics of SO$_2$, since interaction of 2\,$^1$A$_1$ with the dissociative state gives rise to an avoided crossing with the $\mathrm{\tilde{C}}$ state, causing it to correlate adiabatically to the ground state $\mathrm{SO}(^3\Sigma^-)+\mathrm{O}(^3\mathrm{P})$ product channel. \cite{19912792,Becker_SO2_phofex,Katagiri_SO2_photodissoc,ButlerSO2Emission,Nachtigall1999441,Parsons2000499,Bludsk2000607,SO2_dissoc__SOvibdist}

We have also developed a model to explain quantitatively the increasing effective barrier height as a function of bending quantum number, $v_2$. As quanta of $v_2$ are added, the effective bond angle increases and the geometry approaches that of the conical intersection with the 2\,$^1$A$_1$ state, calculated to occur at $\sim$145--150$^{\circ}$. The model quantitatively reproduces the observed increase in level staggering of the $v_3$ progression as a function of $v_2$ ($\sim$5 cm$^{-1}$ per quantum of $v_2$). Our work provides information against which to compare future \textit{ab initio} calculations of the vibronic coupling around the equilibrium geometry of the $\mathrm{\tilde{C}}$ state. 

Finally, our work demonstrates the ability of high-resolution spectroscopy on comparatively simple, low-lying vibrational energy levels to provide useful qualitative information about interactions that occur at much higher energies. The relative simplicity of these low-lying vibrational levels provides an advantage over spectroscopic experiments at higher energy, where assignments are often ambiguous if not impossible. We have used the low-lying vibrational structure in the $\mathrm{\tilde{C}}$ state of SO$_2$ to identify signatures of a three-state vibronic interaction mechanism, as well as the approach toward a conical intersection along the bending coordinate. 


\section{Acknowledgments} \label{Acknowledgments}
The authors thank Anthony Merer and John Stanton for valuable discussions. This material is based upon work supported by the U.S. Department of Energy, Office of Science, Chemical Sciences Geosciences and Biosciences Division of the Basic Energy Sciences Office, under Award Number DE-FG0287ER13671. 

\bibliographystyle{unsrt}
\bibliography{IRUV_Library}

\end{document}